\documentstyle[12pt,A4]{article}

\begin{document}
\hbadness=10000
\begin{titlepage}
\nopagebreak
\vspace*{1.5 cm}
\begin{flushright}

        {\normalsize
 Kanazawa-94-02\\
 KITP-9401\\

January, 1994   }\\
\end{flushright}
\vspace{2cm}
\vfill
\begin{center}
\renewcommand{\thefootnote}{\fnsymbol{footnote}}
{\large \bf Minimal String Model in $Z_4$, $Z_6$ and $Z_8$ Orbifold
 Constructions }

\vfill
\vspace{1.1cm}

{\bf Hiroshi Kawabe, Tatsuo Kobayashi } and

{\bf Noriyasu Ohtsubo$^*$}\\

\vspace{1.1cm}
       Department of Physics, Kanazawa University, \\
       Kanazawa, 920-11, Japan \\
 and \\
$^*$Kanazawa Institute of Technology, \\
Ishikawa 921, Japan

\vfill
\end{center}
\vspace{1.1cm}

\vfill
\nopagebreak
\begin{abstract}
We examine whether a minimal string model possessing the same massless spectra
 as the MSSM can be obtained from $Z_4$, $Z_6$ and $Z_8$ orbifold
 constructions.
Using an anomaly cancellation condition of the target space duality symmetry,
 we derive allowable values of a level $k_1$ of U(1)$_Y$ for the minimal
 string model on the orbifolds through computer analyses.
We investigate threshold corrections of the gauge coupling constants of SU(3),
 SU(2) and U(1)$_Y$ and examine consistencies of the model with the LEP
 experiments.
It is found that $Z_4$ and $Z_8$-II can not derive the minimal string model
 but $Z_6$-I, $Z_6$-II and $Z_8$-I are possible to derive it with $k_1=29/21$,
 $1\leq k_1\leq 32/21$ and $1\leq k_1\leq 41/21$ respectively.
The minimum values of the moduli on unrotated planes are estimated within the
 ranges of the levels.

\end{abstract}

\vfill
\end{titlepage}
\pagestyle{plain}
\newpage
\voffset = 0.5 cm

\leftline{\large \bf 1. Introduction}
\vspace{0.8 cm}

An orbifold compactification is one of the most attractive procedures deriving
 a 4-dim unified theory from superstring theories \cite{ZNOrbi}.
Much work has been devoted in order to construct the minimal supersymmetric
 standard model (MSSM) from the orbifold compactification and to study their
 phenomenological aspects \cite{ZNOrbi2,Anom}.
The superstring theories imply that gauge coupling constants are unified at
 a string scale $M_{\rm st}=5.27\times g_{\rm st}\times 10^{17}$ GeV
 \cite{Kaplunovsky,Derendinger}, where $g_{\rm st}\simeq 1/\sqrt 2$ is
 the universal string coupling constant.
On the other hand, recent LEP measurements support that gauge coupling
 constants  of the standard gauge group ${\rm SU(3)\times SU(2)\times U(1)}_Y$
 are unified at $M_X\simeq 10^{16.2}$ GeV within the framework of the MSSM
 \cite{MSSM} with a Kac-Moody level $k_1=5/3$ of ${\rm U(1)}_Y$ proposed by
 GUTs.
It seems that this difference between $M_X$ and $M_{\rm st}$
 rejects the possibility for a minimal string model which has the same massless
 spectra as the MSSM.

It has been pointed out that the difference between $M_X$ and $M_{\rm st}$
 may be explained by threshold effects due to towers of higher massive modes.
The threshold corrections have been calculated in the case of the orbifold
 models \cite{Kaplunovsky,Derendinger,Dixon}.
In the calculation a target-space duality symmetry \cite{Kikkawa} plays an
 important role.
In ref.~\cite{Ibanez}, the unification of the gauge coupling constants of
 SU(3), SU(2) and U(1)$_Y$ was studied with considering the threshold effects.
It was shown that the minimal string model can not be derived from $Z_3$,
 $Z_7$  and $Z_2\times Z_2$ orbifold models but may be derived from $Z_6$-II,
 $Z_8$-I and the remainig $Z_N\times Z_M$ orbifold models \cite{ZNM} without
 conflicting the unification of the coupling constants
 in the case of $k_1=5/3$.
There is no reason, however, why we choose $k_1=5/3$ in the minimal string
 model, where the level is arbitrary \cite{Ibanez2}.

For the $Z_8$-I, an explicit research for the minimal string model was studied
 without such a restriction of $k_1$ in ref.~\cite{KKO}.
For the $Z_N\times Z_M$ orbifolds, ranges of $k_1$ consistent with the LEP
 experiments were derived in ref.~\cite{KKO2}.

In this paper, we study the minimal string model on $Z_4$, $Z_6$-I, $Z_6$-II,
 $Z_8$-I and $Z_8$-II orbifolds and derive ranges of the levels which lead
 threshold corrections consistent with measured values of the gauge coupling
 constants.
In the next section, we briefly review the $Z_N$ orbifold models and
 discuss massless conditions which constrain oscillator numbers and the levels.
In section three, we review the duality symmetry and the threshold corrections
 to the gauge coupling constants.
Then we study the unification of the SU(3) and SU(2) gauge coupling constants
 in the minimal string model.
In section four, we examine through computer analyses whether the minimal
 string model can be derived from the $Z_4$, $Z_6$-I, $Z_6$-II, $Z_8$-I and
 $Z_8$-II orbifolds.
Then we estimate the ranges of the allowable levels of U(1)$_Y$ and minimum
 values of the moduli.
The last section is devoted to the conclusions.

\vspace{0.8 cm}
\leftline{\large \bf 2. $Z_N$ Orbifold Models}
\vspace{0.8 cm}

In the orbifold models, the string states consist of the bosonic strings on the
 4-dim space-time and a 6-dim orbifold, their right-moving superpartners and
 left-moving gauge parts whose momenta span a shifted $E_8 \times E'_8$
 lattice.
The right-moving fermionic parts are bosonized and momenta of the bosonized
 fields span an SO(10) lattice.
The 6-dim orbifolds are obtained through the division of a 6-dim space R$^6$
 by 6-dim Lie lattices and their automorphisms (twists).
The Lie lattices of our concern are listed in the second column of Table 1
 \cite{Moduli}.
We denote eigenvalues of the twist $\theta$ in a complex basis
 ($X_i,\tilde X_i$) ($i=1,2,3$) as exp$[2\pi i v^i]$
 respectively, whose exponents $v_i$ are exhibited in the third
 column of Table 1.
The twist $\theta$ is embedded into the SO(10) and
 $E_8 \times E'_8$ lattices in terms of shifts so that the $N=1$ SUSY remains
 and the gauge group breaks into a small one.
The $E_8 \times E'_8$ lattice is shifted by Wilson lines \cite{Moduli,WL},
 as well.

There are two types of closed strings on the orbifolds.
One is an untwisted string whose massless states should satisfy
$$h-1=0, \eqno(2.1)$$
where $h$ is a conformal dimension of the $E_8\times E'_8$ gauge part.
The other is a twisted string.
Massless states of $\theta^\ell$-twisted sector should satisfy
 the following condition:
$$h+N_{\rm OSC}+c_{\ell}-1=0, \eqno(2.2)$$
where $N_{\rm OSC}$ is an oscillation number and $c_{\ell}$ is obtained from
$$ c_{\ell}= {1\over 2}\sum_{i=1}^3 v^i_{\ell}(1-v^i_{\ell}),\qquad v^i_{\ell}
\equiv \ell v^i-{\rm Int}(\ell v^i).
\eqno(2.3)$$
Here ${\rm Int}(a)$ represents an integer part of $a$.

A representation $\underline{R}$ of the non-abelian group $G$ contributes to
 the conformal dimension as
$$ h={C(\underline{R}) \over C(G)+k},\eqno(2.4)$$
where $k$ is a level of a Kac-Moody algebra corresponding to $G$ and
 $C(\underline{R})$ ($C(G)$) is a quadratic Casimir of the $\underline{R}$
 (adjoint) representation, e.g., $C({\rm SU}(N))=N$.
In general the string theories derive the gauge groups with $k=1$,
\footnote{Gauge groups with $k\not=1$ are discussed in ref.~\cite{k}}
 except for U(1).
Then we restrict ourselves to the case where $k=1$ for SU(3) and SU(2).
In the minimal string model, the level $k_1$ of U(1)$_Y$ is a
 positive free parameter.

The representations (3,2), ($\bar 3,1$) and (1,2) of the
 ${\rm SU(3) \times SU(2)}$ group have the conformal dimensions of
 $h=7/12,$ 1/3 and 1/4, respectively.
A state with a charge $Q$ of the U(1)$_Y$ group has an additional contribution
 of $h=Q^2/k_1$.
{}From eq.~(2.1) we find that $(3,2)_{1/6}$ ($Q=1/6$) in the untwisted sector
 should satisfy
$$ {7\over 12}+{1\over 36k_1}-1\leq 0. \eqno{(2.5)} $$
Then we have a restriction $k_1\geq 1/15$ in order to obtain $(3,2)_{1/6}$
 in the untwisted sector.
Similarly, we get restrictions for the other representations in the untwisted
 sector, as shown in Table 2.

{}From eq.~(2.2) we find that $(3,2)_{1/6}$ in the $\theta^{\ell}$-twisted
sector
 has oscillators $N_{\rm OSC}\le 5/12-c_{\ell}$ under a condition
 $k_1 \ge 1/(15-36N_{\rm OSC}-36c_{\ell})$.
Similarly we can obtain the allowable values of $N_{\rm OSC}$ and conditions
 of $k_1$ for the other representations.

\vspace{0.8 cm}
\leftline{\large \bf 3. Duality and Threshold Corrections}
\vspace{0.8 cm}

It is plausible that the duality symmmetry is retained in effective field
 theories derived from the orbifold models.
In the theories, moduli fields $T_i$ ($i=1,2,3$) associated with the $i$-th
 complex planes have the K\"ahler potentials
$$ -\sum_i{\rm log}|T_i+\bar T_i|,\eqno(3.1)$$
which are invariant under a duality transformation:
$$ T_i \rightarrow {a_iT_i-ib_i \over ic_iT_i+d_i} ,\eqno(3.2)$$
up to the K\"ahler transformation, where $a_i,b_i,c_i,d_i\in {\bf Z}$ and
 $a_id_i-b_ic_i=1$.

The K\"ahler potential of the matter field $A$ is
$$ \prod^3_{i=1}(T_i+\bar T_i)^{n^i}A\bar A,\eqno(3.3)$$
whose duality invariance requires the following transformation:
$$ A \rightarrow A \prod_{i=1}^3(ic_iT_i+d_i)^{n^i},\eqno(3.4)$$
where $n^i$ is called a modular weight \cite{Dixon2,Ibanez}.

For the untwisted sector associated with the $p$-th plane, the matter fields
 have $n^i=-\delta^i_p$, as shown in Table 2, where an underline represents
 any permutation of the elements.
The $\theta^\ell$-twisted state without oscillators has the following
 modular weights:
$$\begin{array}{llllll}
n^i&=&1-v^i_{\ell}, & \qquad v^i_{\ell}& \neq & 0, \\
n^i&=&0,            & \qquad v^i_{\ell}& =    & 0.
\end{array}
\eqno(3.5)$$
The oscillator $\partial X_i$ reduces the corresponding element of the modular
 weight by one and the oscillator $\partial \tilde X_i$ contributes oppositely.
Thus we can obtain modular weights of the matter fields using the allowable
 values for $N_{\rm OSC}$.

Table 3 lists the modular weights and the lower bounds of $k_1$ of each
 representations permitted by the massless condition in the previous section
 for $\theta$- and $\theta^2$-twisted sectors in the $Z_4$ orbifold model.
(In the table $\theta^3$-twisted sector is omitted because it includes only
 anti-matters \cite{Yukawa}.
Such sectors are also omitted in the following tables.)
For example, the representations $(3,2)_{1/6}$, $(\bar 3,1)_{1/3}$,
 $(\bar 3,1)_{-2/3}$, $(1,2)_{\pm 1/2}$ and $(1,1)_1$ in $\theta$ twisted
 sector are able to possess $n^i=(-3,-3,-2)/4$, if $k_1\geq 4/15$, $16/51$,
 $64/51$, $4/7$ and $16/11$ respectively.
The modular weight $n^i=(\underline{-7,-3},-2)/4$ is impossible for
 $(3,2)_{1/6}$, and  $(\underline{-11,-3},-2)/4$, $(-7,-7,-2)/4$,
 $(-3,-3,-6)/4$ and $(-3,-3, 2)/4$ are also impossible for $(3,2)_{1/6}$,
 $(\bar 3,1)_{1/3}$, $(\bar 3,1)_{-2/3}$ and $(1,2)_{\pm 1/2}$.
In the table, parentheses of the lower bounds of $k_1$ represent the values
 exceed 2.
It seems that the orbifold models compatible with the experiments derive
 $k_1 < 2$ as suggested in the previous study about the $Z_N\times Z_M$
 orbifolds \cite{KKO2}.
Then we limit the subsequent studies to the cases of $k_1<2$.
For the twisted sectors of $Z_6$-I, $Z_6$-II, $Z_8$-I and $Z_8$-II, the
 modular weights and the lower bounds of $k_1$  are listed in Table 4, Table
 5, Table 6 and Table 7, respectively, where we omit the modular weights
 requiring $k_1 \geq 2$.

The duality symmetry becomes anomalous by loop effects.
Anomaly coefficients $b'^i_a$, ($a=3$, 2 and 1 correspond to SU(3), SU(2) and
 U(1) respectively) are given by
$$b'^i_a=-C(G_a)+\sum_{\underline R} T(\underline R)(1+2n^i_{\underline R}),
\eqno(3.6)$$
where $T(\underline R)$ is an index given by
 $T(\underline R)=C(\underline R){\rm dim}(\underline R)/{\rm dim}(G)$, e.g.,
 $T(\underline R)=1/2$ for the $N$-dim fundamental representation of SU($N$).
The anomaly can be cancelled only by the GS mechanism, if the following
 condition is satisfied:
$$b'^j_3=b'^j_2=b'^j_1/k_1, \eqno(3.7)$$
for $j$-th planes rotated under any twist.\footnote{Although the anomaly
 coefficients of the hidden sector must be considered in eq.~(3.7), we
 disregard them in this paper.
The hidden sector in $Z_8$-I orbifold was investigated in ref.~\cite{koba}
 with respect to the anomaly cancellation and SUSY breaking.}
Only the first plane is concerned with eq.~(3.7) in $Z_6$-II, whereas
 the first and the second planes are in $Z_4$, $Z_6$-I, $Z_8$-I and $Z_8$-II.
The equation (3.7) gives more stringent constraints to the latter orbifolds
 because $k_1$ is common for the both planes.
The $Z_N\times Z_M$ orbifold models are free from the anomaly cancellation
 condition (3.7), because the orbifolds do not have such complex planes
 \cite{Ibanez,KKO2}.

The other $k$-th ($k\not= j$) planes contribute in the threshold corrections
 of the gauge coupling constants induced by the tower of higher massive modes.
The threshold corrections \cite{Derendinger,Dixon} are given by
$$\Delta_a(T_k)=-{1 \over 16\pi^2}\sum_k (b'^k_a-k_a\delta^k_{\rm GS})
{\rm log}|\eta(T_k)|^4, \eqno(3.8)$$
where $\delta_{\rm GS}$ is a Green-Schwarz coefficient \cite{GS} independent
 of the gauge groups and $\eta(T)=e^{-\pi T/12}\prod_{n\ge 1}(1-e^{-2\pi nT})$
 is the Dedekind function.
Eq.~(3.8) may be modified in the cases of other lattices in Table 1
 \cite{Mayr}.
The moduli $T_2$ and $T_3$ participate in eq.~(3.8) for the $Z_6$-II orbifold,
 while only $T_3$ for the other orbifolds.
For simplicity, we consider a case of $T_2 = T_3(=T_k)$ in $Z_6$-II.
Then we denote $\delta^k_{\rm GS}=\delta^2_{\rm GS}+\delta^3_{\rm GS}$,
 $b'^k_a=b'^3_a+b'^2_a$ and $n^k_{\underline R}=n^2_{\underline R}
+n^3_{\underline R}$ for $Z_6$-II and remove the summation.

Using the threshold corrections, we obtain the one-loop coupling constants
 $\alpha_a(\mu)=k_ag_a^2(\mu)/4\pi$ ($k_3=k_2=1$) at an energy scale $\mu$ as
 follows,
$$ \alpha^{-1}_a(\mu)=\alpha^{-1}_{\rm st}+{1 \over 4\pi}{b_a\over k_a}
 {\rm log}{M_{\rm st}^2 \over \mu^2}-{1\over 4\pi}
 (b'^k_a-k_a\delta^k_{\rm GS}){\rm log}[(T_k+\bar T_k)|\eta(T_k)|^4],
 \eqno(3.9)$$
where $\alpha_{\rm st}=g^2_{\rm st}/4\pi$ and $b_a$ are $N=1$
 $\beta$-function coefficients.
We use the same $b_3=-3$, $b_2=1$ and $b_1=11$ as ones of the MSSM.

{}From this renormalization group flow, we can derive a unified scale $M_X$
from
$$ {\rm log}{M_{X} \over M_{\rm st}}={1\over 8} (b'^k_3-b'^k_2)
 {\rm log}[(T_k+\bar T_k)|\eta(T_k)|^4] ,\eqno(3.10)$$
where  $\alpha^{-1}_3(M_X)=\alpha^{-1}_2(M_X)$.\footnote{When $k_1=5/3$,
 the LEP measurements are consistent with $\alpha^{-1}_3(M_X)=\alpha^{-1}_2
 (M_X)=\alpha^{-1}_1(M_X)$ within the framework of the MSSM.}
Since the LEP measurements indicate $M_X<M_{\rm st}$, eq.~(3.10) gives a
 restriction $\Delta b'^k\equiv b'^k_3-b'^k_2 > 0,$ because
 log$[(T_k+\bar{T_k})|\eta(T_k)|^4]$ is always negative.
When $\Delta b'^k=3$, we get $T_k\simeq 12$ through $M_X\simeq 10^{16.2}$ GeV
 and $M_{\rm st}\simeq 3.73\times 10^{17}$ GeV.
Since such a large $T_k$ is not desirable, we impose $\Delta b'^k\geq 3$ on the
 subsequent analyses.

The one-loop fine structure constant of the electro-magnetic interaction is
 obtained from $\alpha^{-1}_{\rm em}=k_1\alpha^{-1}_1+\alpha^{-1}_2$ as
\renewcommand{\arraystretch}{1.5}
\arraycolsep=0.5mm
$$ \begin{array}{rll}
\alpha^{-1}_{\rm em}(\mu)&=&{\displaystyle (k_1+1)\alpha^{-1}_{\rm st}+{3 \over
 \pi}{\rm log}{M_{\rm st}^2 \over \mu^2}}\\
 &-&{\displaystyle {1\over 4\pi} (b'^k_1+b'^k_2-(k_1+1)
 \delta^k_{\rm GS}){\rm log}[(T_k+\bar T_k)|\eta(T_k)|^4].}
\end{array} \eqno(3.11)$$
By means of $\sin ^2\theta_{\rm W}=\alpha_{\rm em}/\alpha_2$, we derive
\renewcommand{\arraystretch}{1.5}
\arraycolsep=0.5mm
$$ \begin{array}{rl}
\sin ^2\theta_{\rm W}(\mu)={\displaystyle {1\over k_1+1}+{\alpha_{\rm em}
(\mu)\over 4\pi (k_1+1)}}
\bigg\{ &{\displaystyle (k_1-11){\rm log}{M_{\rm st}^2 \over \mu^2}}\\
       -&(k_1b'^k_2-b'^k_1){\rm log}[(T_k+ \bar{T}_k)|\eta(T_k)|^4]\bigg\} .
\end{array}
\eqno(3.12)$$
Making use of eqs.~(3.9), (3.11) and (3.12) so as to remove $\alpha_{\rm st}$,
 $\delta^k_{\rm GS}$ and $T_k$, we obtain
\renewcommand{\arraystretch}{1.5}
\arraycolsep=0.5mm
$$ \begin{array}{rll}
(k_1b'^k_2-b'^k_1)\alpha_3^{-1}(\mu)&=& (b'^k_2-b'^k_3)\alpha_{\rm em}^{-1}
(\mu)-\left\{ b'^k_1+b'^k_2-(k_1+1)b'^k_3\right\} \alpha_{\rm em}^{-1}(\mu)
\sin^2 \theta_{\rm W}(\mu)\\
&+&{\displaystyle {1\over 4\pi}\left\{ 4b'^k_1-(3k_1+11)b'^k_2-(k_1-11)b'^k_3
\right\} {\rm log}{M_{\rm st}^2\over \mu^2}}.
\end{array} \eqno(3.13)$$
Through eq.~(3.13) one can check consistencies of orbifold models with the LEP
experiments as discussed in the next section.

\newpage
\vspace{0.8 cm}
\leftline{\large \bf 4. Level $k_1$ and Modulus $T_k$}
\vspace{0.8 cm}

At first we derive the level $k_1$ of the U(1)$_Y$ from eq.~(3.7).
For concreteness, we exhibit the $Z_6$-I orbifold model.
The anomaly coefficents $b'^j_a$ are obtained from inserting values of
 three $n^j_{(3,2)}$, three $n^j_{(\bar 3,1)_{1/3}}$, three
 $n^j_{(\bar 3,1)_{-2/3}}$, five $n^j_{(2,1)}$ and three $n^j_{(1,1)}$
 permitted in Table 2 and Table 4 into eq.~(3.6).
Through computer analyses we have found 120016 combinations (up to
 combinations of the singlets) of $n^j_{\underline R}$ satisfying the first
 equality of eq.~(3.7) with respect to the complex planes $j=1,2$.
It has been also found that 1658 combinations of them pass the condition
 $\Delta b'^k\geq 3$.
Among them, 247 combinations including the singlets satisfy the full
 equalities of eq.~(3.7) and the condition $\Delta b'^k\geq 3$ simultaneously.
Then we have obtained values $k_1$ of the above models in terms of ratios
 between anomaly coefficients.
The values $k_1$ have been checked whether they are not smaller than the lower
 bounds imposed on the levels for each $n_{\underline R}$.
At last we have examined compatibility with experiments through eq.~(3.13).
Here we consider the case of $M_Z$ SUSY breaking scale and use the
 experimental values $\sin ^2\theta_{\rm W}(M_Z)=0.2325\pm 0.0008$,
 $\alpha_{\rm em}^{-1}(M_Z)=127.9\pm 0.1$ and $\alpha^{-1}_3(M_Z)=
8.82\pm 0.27$ at $M_Z=91.173\pm 0.020$ GeV.
In this way we have obtained only one model, whose level is 29/21.

We show modular weights of the matter fields in the $Z_6$-I model.
The $(3,2)_{1/6}$ representations have modular weights of two $(0,0,-1)$
 and one $(-5,-5,-2)/6$.
The $(\bar 3,1)_{2/3}$ representations have two $(-5,-5,-2)/6$ and
 one $(-4,-4,-4)/6$.
The whole $(\bar 3,1)_{-1/3}$ have $(-5,-5,4)/6$.
Five $(1,2)_{\pm 1/2}$ representations have two $(-11,-5,-2)/6$, two\break
 $(-5,-11,-2)/6$ and one $(-4,-4,-4)/6$.
The $(1,1)_{1}$ representations have two $(0,0,-1)$ and one $(-5,-5,-2)/6$.
In order to verify the existance of this model, one must investigate through
 the similar procedure to ref.~\cite{KKO}.
In this investigation, however, much work can be reduced by the informations
 about the twisted sectors and the oscillators which are obtained from the
 above modular weights by means of Table 2 and Table 4.

Through the similar computer analyses we have found no model for the $Z_4$
 orbifold, because it leads $\alpha^{-1}_3(M_Z)\leq 3.97$ from eq.~(3.13).
The $Z_6$-II orbifold has 4573 models with $1\leq k_1\leq 32/21$.
Unless we impose the condition $\Delta b'^k\geq 3$, we can get 6 models with
 $k=5/3$ as its maximum value.
Since the models require $\Delta b'^k=2$ or 1, their moduli $T_2$ and $T_3$
 must be unnaturally large.
The $Z_8$-I orbifold derives 104 models with $1\leq k_1\leq 41/21$, which
 includes the GUT prediction value $k_1=5/3$ as emphasized in
 ref.~\cite{Ibanez}.
The $Z_8$-II orbifold, however, can not derive the minimal string model
 because no solution of eq.~(3.7) is found within $1\leq k_1< 2$.

The ranges of $k_1$ derived in the above estimations forbid some modular
 weights of the twisted matter fields in Table 4, Table 5, Table 6 and Table 7
 for the $Z_6$-I, $Z_6$-II, $Z_8$-I and $Z_8$-II orbifolds, respectively .
In $Z_6$-I, for instance, $(\bar 3,1)_{-2/3}$ are not prohibitted to possess
 $n^i=(\underline{-11,-5},-2)/6$ on $\theta$-twisted sector for $k_1<2$, but
 for $k_1=29/21$.
These restrictions of the modular weights rule out higher $N_{\rm OSC}$.
Thus, one can reduce extents of the minimal string model searches on the
 orbifolds.

Next, we investigate the moduli $T_k$ of the planes unrotated under some
 twists.
{}From eq.~(3.10) we can get Re$T_k$ when $\Delta b'^k$ is obtained.
The model of the $Z_6$-I have $\Delta b'^3=3$.
Then we get $T_3\simeq 12$.
The maximum value of $\Delta b'$ is 6 (4) for the minimal string models of
 $Z_6$-II ($Z_8$-I).
Then we obtain minimum value of $T_k$ as 7.1 (9.7).
Although the $Z_6$-I must have large $T_k$, the $Z_6$-II and the $Z_8$-I
 are possible to have reasonable values of $T_k$.

\vspace{0.8 cm}
\leftline{\large \bf 5. Conclusions}
\vspace{0.8 cm}

In this paper we have discussed the minimal string model on the
 $Z_4$, $Z_6$-I, $Z_6$-II, $Z_8$-I and $Z_8$-II orbifolds.
We have studied the anomaly cancellation of the target space duality and the
 threshold corrections of the gauge coupling constants.
Using the computer analyses, we have examined whether the orbifolds can derive
 the minimal string model satisfying the anomaly cancellation condition and
 being consistent with the experimental values of $\sin^2 \theta_{\rm W}$,
 $\alpha^{-1}_{\rm em}$, $\alpha^{-1}_3$ and $M_Z$.
It has been found that the $Z_4$ and $Z_8$-II orbifolds can not derive the
 minimal string model.
We have also found that the minimal string models on the $Z_6$-I, $Z_6$-II and
 $Z_8$-I orbifolds are possible to have the levels $k_1=29/21$,
 $1\leq k_1\leq 32/21$ and $1\leq k_1\leq 41/21$
 respectively, under the condition $\Delta b'^k \geq 3$ for removing large
 moduli $T_k$.
Although the GUT prediction value $k_1=5/3$ is included only in $Z_8$-I, it is
 also included in the $Z_6$-II as discussed in ref.~\cite{Ibanez} if $\Delta
 b'^k=2$ or 1 is permitted.

The modulus $T_k$ of the unrotated plane has been
 studied in order to explain the discrepancy between the unified scale
 $M_X$ of $g_3$ and $g_2$ and the string scale $M_{\rm st}$.
We have found that the minimal string models on $Z_6$-I, $Z_6$-II and $Z_8$-I
 orbifolds must have $T_k\geq 12$, 7.1 and 9.7 respectively.

Although we have not investigated $Z_{12}$-I and $Z_{12}$-II orbifold models,
 the above procedure can be also applied to them.
One will be able to investigate the minimal string model with extra matters
 \cite{extra} through the similar estimations.

\vspace{0.8 cm}
\leftline{\large \bf Acknowledgement}
\vspace{0.8 cm}

The authors would like to thank D.~Suematsu for useful discussions.


\newpage
\pagestyle{empty}
\noindent
\begin{center}
{\bf \large Table 1. Restrictive values of $k_1$ and $T$}

\vspace{10mm}
\renewcommand{\arraystretch}{1.5}

\begin{tabular}{|c|c|c||c|c|c|}
\hline
Orbifold & Lattice                  &  $v_i$    &  $k_1$
   & $\Delta b'^k$ &  $T_k$
   \\ \hline \hline
$Z_4$    & SO(5)$^2\times$SU(2)$^2$ & (1,1,2)/4 &     -          &    -   & -
   \\ \hline
$Z_6$-I  & $G_2^2\times$SU(3)       & (1,1,4)/6 &  $29/21$
  & $\leq 3$ & $\geq 12$   \\ \hline
$Z_6$-II & $G_2\times$SU(3)$\times$SU(2)$^2$ & (1,2,3)/6 & $1\sim 32/21$
  & $\leq 6$   & $\geq 7.1$    \\ \hline
$Z_8$-I  & SO(9)$\times$SO(5)       & (1,5,2)/8 &  $1\sim 41/21$
  & $\leq 4$ & $\geq 9.7$    \\ \hline
$Z_8$-II & SO(9)$\times$SU(2)$^2$   & (1,3,4)/8 &     -          &    -   & -
   \\ \hline
                                   \end{tabular}

\end{center}
\vspace{30mm}
\begin{center}
{\bf \large Table 2. Modular weights in untwisted sector}
\vspace{10mm}

\renewcommand{\arraystretch}{1.5}
\begin{tabular}{|c|c|c|c|c|c|}
\hline
  Modular weight & \multicolumn{5}{c|}{Lower-bound of $k_1$} \\ \cline{2-6}
  $n^i$   & $(3,2)_{1/6}$ & $(\bar 3,1)_{1/3}$ & $(\bar 3,1)_{-2/3}$
 & $(1,2)_{\pm 1/2}$ & $(1,1)_1$  \\ \hline \hline
  $(\underline{-1,0,0})$  &   1/15  &      1/6      &      2/3      &   1/3
 & 1   \\ \hline
                                   \end{tabular}

\end{center}
\newpage
\noindent
\begin{center}
{\bf \large Table 3. modular weights in twisted sectors on $Z_4$ orbifold}
\vspace{10mm}

\renewcommand{\arraystretch}{1.5}
\begin{tabular}{|c||c|c|c|c|c|c|c|}
\hline
 Twisted  & modular weight & \multicolumn{5}{c|}{Lower-bound of $k_1$} \\
\cline{3-7}
 sector   & $n^i$   & $(3,2)_{1/6}$ & $(\bar 3,1)_{1/3}$
 & $(\bar 3,1)_{-2/3}$ & $(1,2)_{\pm 1/2}$ & $(1,1)_1$  \\ \hline \hline
$\theta$  & $(-3,-3,-2)/4$           & 4/15 & 16/51 &  64/51 & 4/7  & 16/11 \\
          & $(\underline{-7,-3},-2)/4$  & - & 16/15 & (64/51)& 4/3  &(16/7) \\
          & $(\underline{-11,-3},-2)/4$ & - &  -    &    -   &  -   &(16/3) \\
          & $(-7,-7,-2)/4$              & - &  -    &    -   &  -   &(16/3) \\
          & $(-3,-3,-6)/4$              & - &  -    &    -   &  -   &(16/3) \\
          & $(-3,-3, 2)/4$              & - &  -    &    -   &  -   &(16/3) \\
 \hline
$\theta^2$& $(-2,-2,0)/4$            & 1/6  & 4/15  &  16/15 & 1/2  & 4/3  \\
          & $(\underline{-6,-2},0)/4$   & - &  -    &    -   &  -   &(4)  \\
          & $(\underline{2,-2},0)/4$    & - &  -    &    -   &  -   &(4)  \\
 \hline
                                   \end{tabular}

\end{center}
\newpage
\begin{center}
\noindent

\begin{center}
{\bf \large Table 4. Modular weights in twisted sectors on $Z_6$-I orbifold}
\vspace{10mm}

\renewcommand{\arraystretch}{1.5}
\begin{tabular}{|c||c|c|c|c|c|c|c|}
\hline
 Twisted  & modular weight & \multicolumn{5}{c|}{Lower-bound of $k_1$} \\
\cline{3-7}
 sector   & $n^i$   & $(3,2)_{1/6}$ & $(\bar 3,1)_{1/3}$
 & $(\bar 3,1)_{-2/3}$ & $(1,2)_{\pm 1/2}$ & $(1,1)_1$  \\ \hline \hline
$\theta$  & $(-5,-5,-2)/6$           & 1/6  & 4/15  &  16/15 & 1/2  & 4/3   \\
          & $(\underline{-11,-5},-2)/6$  & - & 4/9  &  16/9  & 3/4  & 12/7  \\
          & $(\underline{-17,-5},-2)/6$ & - &  4/3  &    -   & 3/2  &   -   \\
          & $(-11,-11,-2)/6$            & - &  4/3  &    -   & 3/2  &   -   \\
          & $(-5,-5,4)/6$               & - &  4/3  &    -   & 3/2  &   -   \\
 \hline
$\theta^2$& $(-4,-4,-4)/6$           & 1/3  & 1/3   &  4/3   & 3/5  & 3/2  \\
 \hline
$\theta^3$& $(-3,-3,0)/6$            & 1/6  & 4/15  & 16/15  & 1/2  & 4/3  \\
 \hline
                                   \end{tabular}

\end{center}
\newpage
\begin{center}
\noindent

\begin{center}
{\bf \large Table 5. Modular weights in twisted sectors on $Z_6$-II orbifold}
\vspace{10mm}

\renewcommand{\arraystretch}{1.5}
\begin{tabular}{|c||c|c|c|c|c|c|c|}
\hline
 Twisted  & modular weight & \multicolumn{5}{c|}{Lower-bound of $k_1$} \\
\cline{3-7}
 sector   & $n^i$   & $(3,2)_{1/6}$ & $(\bar 3,1)_{1/3}$
 & $(\bar 3,1)_{-2/3}$ & $(1,2)_{\pm 1/2}$ & $(1,1)_1$  \\ \hline \hline
$\theta$  & $(-5,-4,-3)/6$ & 1/3  & 4/13  &  16/13  & 9/16 & 36/25 \\
          & $(-11,-4,-3)/6$ & -   & 4/7  &     -    & 9/10 & 36/19 \\
 \hline
$\theta^2$& $(-4,-2,0)/6$  & 1/7  & 1/5   &  4/5    & 9/19 & 9/7  \\
          & $(-10,-2,0)/6$ &  -   & 1/2  &   2      & 9/7  &  -   \\
          & $(-4,4,0)/6$   &  -   & 1/2  &   2      & 9/7  &  -   \\
 \hline
$\theta^3$& $(-3,0,-3)/6$            & 1/6  & 4/15  & 16/15  & 1/2  & 4/3  \\
 \hline
$\theta^4$& $(-2,-4,0)/6$  & 1/7  & 1/5   &  4/5    & 9/19 & 9/7  \\
          & $(-2,-10,0)/6$ &  -   & 1/2  &   2      & 9/7  &  -   \\
          & $(4,-4,0)/6$   &  -   & 1/2  &   2      & 9/7  &  -   \\
 \hline
                                   \end{tabular}

\end{center}
\newpage
\begin{center}
\noindent

\begin{center}
{\bf \large Table 6. Modular weights in twisted sectors on $Z_8$-I orbifold}
\vspace{10mm}

\renewcommand{\arraystretch}{1.5}
\begin{tabular}{|c||c|c|c|c|c|c|c|}
\hline
 Twisted  & modular weight & \multicolumn{5}{c|}{Lower-bound of $k_1$} \\
\cline{3-7}
 sector   & $n^i$   & $(3,2)_{1/6}$ & $(\bar 3,1)_{1/3}$
 & $(\bar 3,1)_{-2/3}$ & $(1,2)_{\pm 1/2}$ & $(1,1)_1$  \\ \hline \hline
$\theta$  & $(-7,-3,-6)/8$  & 16/87 & 64/231 & 256/231 & 16/31 & 64/47 \\
          & $(-15,-3,-6)/8$ & 16/15 & 64/159 & 256/159 & 16/23 & 64/39 \\
          & $(-23,-3,-6)/8$ &   -   & 64/87  &    -    & 16/15 &   -   \\
          & $(-7,-3,-14)/8$ &   -   & 64/87  &    -    & 16/15 &   -   \\
 \hline
$\theta^2$& $(-6,-6,-4)/8$  & 4/15  & 16/51  &  64/51  & 4/7 & 16/11  \\
          & $(\underline{-14,-6},-4)/8$ &
                               -    & 16/15  &       -  & 4/3  &    -   \\
 \hline
$\theta^4$& $(-4,-4,0)/8$            & 1/6  & 4/15  & 16/15  & 1/2  & 4/3  \\
 \hline
$\theta^5$& $(-3,-7,-6)/8$  & 16/87 & 64/231 & 256/231 & 16/31 & 64/47 \\
          & $(-3,-15,-6)/8$ & 16/15 & 64/159 & 256/159 & 16/23 & 64/39 \\
          & $(-3,-23,-6)/8$ &   -   & 64/87  &    -    & 16/15 &   -   \\
          & $(-3,-7,-14)/8$ &   -   & 64/87  &    -    & 16/15 &   -   \\
 \hline
                                   \end{tabular}

\end{center}
\newpage
\begin{center}
\noindent

\begin{center}
{\bf \large Table 7. Modular weights in twisted sectors on $Z_8$-II orbifold}
\vspace{10mm}

\renewcommand{\arraystretch}{1.5}
\begin{tabular}{|c||c|c|c|c|c|c|c|}
\hline
 Twisted  & modular weight & \multicolumn{5}{c|}{Lower-bound of $k_1$} \\
\cline{3-7}
 sector   & $n^i$   & $(3,2)_{1/6}$ & $(\bar 3,1)_{1/3}$
 & $(\bar 3,1)_{-2/3}$ & $(1,2)_{\pm 1/2}$ & $(1,1)_1$  \\ \hline \hline
$\theta$  & $(-7,-5,-4)/8$  & 16/69 & 64/213 & 256/213 & 16/29 & 64/45 \\
          & $(-15,-5,-4)/8$ &   -   & 64/141 & 256/141 & 16/21 & 64/37 \\
          & $(-23,-5,-4)/8$ &   -   & 64/69  &    -    & 16/13 &   -   \\
 \hline
$\theta^2$& $(-6,-2,0)/8$  & 4/33  & 16/69  &  64/69  & 4/9 & 16/13  \\
          & $(-14,-2,0)/8$ &   -   & 16/33  &  64/33  & 4/5 & 16/9   \\
          & $(-6,6,0)/8$   &   -   & 16/33  &  64/33  & 4/5 & 16/9   \\
 \hline
$\theta^3$& $(-5,-7,-4)/8$  & 16/69 & 64/213 & 256/213 & 16/29 & 64/45 \\
          & $(-5,-15,-4)/8$ &   -   & 64/141 & 256/141 & 16/21 & 64/37 \\
          & $(-5,-23,-4)/8$ &   -   & 64/69  &    -    & 16/13 &   -   \\
 \hline
$\theta^4$& $(-4,-4,0)/8$            & 1/6  & 4/15  & 16/15  & 1/2  & 4/3  \\
 \hline
$\theta^6$& $(-2,-6,0)/8$  & 4/33  & 16/69  &  64/69  & 4/9 & 16/13  \\
          & $(-2,-14,0)/8$ &   -   & 16/33  &  64/33  & 4/5 & 16/9   \\
          & $(6,-6,0)/8$   &   -   & 16/33  &  64/33  & 4/5 & 16/9   \\
 \hline
                                   \end{tabular}

\end{center}

\newpage
\begin{thebibliography}{99}

\bibitem{ZNOrbi}
L.~Dixon, J.~Harvey, C.~Vafa and E.~Witten, Nucl.~Phys. {\bf B261} (1985)
 678; Nucl.~Phys. {\bf B274} (1986) 285.

L.E.~Ib\'a\~nez, J.~Mas, H.P.~Nilles and F.~Quevedo, Nucl.~Phys. {\bf B301}
 (1988) 157.

\bibitem{ZNOrbi2}
L.E.~Ib\'a\~nez, J.E.~Kim, H.P.~Nilles and F.~Quevedo, Phys.Lett. {\bf B191}
 (1987) 282.

A.~Font, L.E.~Ib\'a\~nez, H.-P.~Nilles and F.~Quevedo, Phys.~Lett. {\bf B210}
 (1988) 101.

J.A~Casas and C.~Mu\~noz, Phys.~Lett. {\bf B209} (1988) 214; Phys.~Lett.
 {\bf B212} (1988) 343; Phys.~Lett. {\bf B214} (1988) 63.

Y.~Katsuki, Y.~Kawamura, T.~Kobayashi, N.~Ohtsubo, Y.~Ono and K.~Tanioka,
 Nucl.~Phys. {\bf B341} (1990) 611.

J.A.~Casas, A.~de~la~Macorra, M.~Mondrag\'on and C.~Mu\~noz, Phys.~Lett.
 {\bf B247} (1990) 50.

\bibitem{Anom}
A.~Font, L.E.~Ib\'a\~nez, H.-P.~Nilles and F.~Quevedo, Nucl.~Phys. {\bf B307}
 (1988) 109.

J.A~Casas, E.K.~Katehou and C.~Mu\~noz, Nucl.~Phys. {\bf B317} (1989) 171.

\bibitem{Kaplunovsky}
V.S.~Kaplunovsky, Nucl.~Phys. {\bf B307} (1988) 145.

\bibitem{Derendinger}
J.-P.~Derendinger, S.~Ferrara, C.~Kounnas and F.~Zwirner, Nucl.~Phys.
 {\bf B372} (1992) 145.

\bibitem{MSSM}
J.~Ellis, S.~Kelley and D.V.~Nanopoulous, Phys.~Lett. {\bf B260} (1991) 131.

U.~Amaldi, W.~de~Boer and H.~F\"urstenau, Phys.~Lett. {\bf B260} (1991) 447.

P.~Langacker and M.~Luo, Phys.~Rev. {\bf D44} (1991) 817.

G.G.~Ross and R.G.~Roberts, Nucl.~Phys. {\bf B377} (1992) 571.

\bibitem{Dixon}
L.J.~Dixon, V.S.~Kaplunovsky and J.~Louis, Nucl.~Phys. {\bf B355} (1991) 649.

I.~Antoniadis, K.S.~Narain and T.R.~Taylor, Phys.~Lett. {\bf B267} (1991) 37.

\bibitem{Kikkawa}
K.~Kikkawa and M.~Yamasaki, Phys.~Lett. {\bf B149} (1984) 357.

N.~Sakai and I.~Senda, Prog.~Theor.~Phys. {\bf 75} (1986) 692.

\bibitem{Ibanez}
L.E.~Ib\'a\~nez and D.~L\"ust, Nucl.~Phys. {\bf B382} (1992) 305.

\bibitem{ZNM}
A.~Font, L.E.~Ib\'a\~nez and F.~Quevedo, Phys.~Lett. {\bf B217} (1989) 272.

T.~Kobayashi and N.~Ohtsubo, Phys.~Lett. {\bf B262} (1991) 425.

\bibitem{Ibanez2}
L.E.~Ib\'a\~nez, Phys.~Lett. {\bf B318} (1993) 73.

\bibitem{KKO}
H.~Kawabe, T.~Kobayashi and N.~Ohtsubo, Preprint Kanazawa-93-08, KITP-9302
 (hep-th/9309069) to be published in Phys. Lett. B.

\bibitem{KKO2}

H.~Kawabe, T.~Kobayashi and N.~Ohtsubo, Preprint Kanazawa-93-12, KITP-9303
 (hep-th/9311166) to be published in Phys. Lett. B.

\bibitem{Moduli}

T.~Kobayashi and N.~Ohtsubo, Int.J.~Mod.~Phys.~{\bf A9} (1994) 87.

\bibitem{WL}
L.E.~Ib\'a\~nez, H.P.~Nilles and F.~Quevedo, Phys.~Lett. {\bf B187} (1987) 25.

T.~Kobayashi and N.~Ohtsubo, Phys.~Lett. {\bf B257} (1991) 56.

\bibitem{k}
A.~Font, L.E.~Ib\'a\~nez and F.~Quevedo, Nucl.~Phys. {\bf B345} (1990) 389.

\bibitem{Dixon2}
L.J.~Dixon, V.S.~Kaplunovsky and J.~Louis, Nucl.~Phys. {\bf B329} (1990) 27.

\bibitem{Yukawa}
T.~Kobayashi and N.~Ohtsubo, Phys.~Lett. {\bf B245} (1990) 441.

\bibitem{koba}
T.~Kobayashi, Preprint Kanazawa-93-11 (hep-th/9310023).

\bibitem{GS}
M.B.~Green and J.H.~Schwarz, Phys.~Lett. {\bf B149} (1984) 117.

\bibitem{Mayr}
P.~Mayr and S.~Stieberger, Preprint MPI-Ph/93-07, TUM-TH-152/93
 (hep-th/9303017)

D.~Bailin, A.Love, W.A.~Sabra and S.~Thomas, Preprint QMW-TH-93/22,
 SUSX-TH-93/14 (hep-th/9310008)

\bibitem{extra}
I.~Antoniadis, J.~Ellis, S.~Kelley and D.V.~Nanopoulos, Phys.~Lett. {\bf B272}
 (1991) 31.

A.E.~Faraggi, Phys.~Lett. {\bf B302} (1993) 202.

\end{thebibliography}
\end{document}